\documentclass[prb,aps, reprint,superscriptaddress,preprintnumbers, 
]{revtex4-1}
\usepackage{graphicx}
\usepackage{dcolumn}
\usepackage{bm}
\usepackage{color}
\usepackage{sidecap}
\usepackage{amssymb}

\begin{document}

\title {Robust random number generation \\using steady-state emission of gain-switched laser diodes}

\author{Z. L. Yuan}
\email{zhiliang.yuan@crl.toshiba.co.uk}
\author{M. Lucamarini}
\author {J. F. Dynes}
\author {B. Fr\"ohlich}
\author{A. Plews}
\author {A.~J.~Shields}
\affiliation{Toshiba Research Europe Limited, Cambridge Research Laboratory, 208 Cambridge Science Park, Milton Road, Cambridge, CB4~0GZ, United Kingdom}
\date{\today}

\begin{abstract}
We demonstrate robust, high-speed random number generation using interference of the steady-state emission of guaranteed random phases, obtained through gain-switching a semiconductor laser diode.
Steady-state emission tolerates large temporal pulse misalignments and therefore significantly improves the interference quality.
Using an 8-bit digitizer followed by a finite-impulse-response unbiasing algorithm, we achieve random number generation rates of 8 and 20~Gb/s, for laser repetition rates of 1 and 2.5~GHz, respectively, with a $\pm$20\% tolerance in the interferometer differential delay. We also report a generation rate of 80~Gb/s using partially phase-correlated short pulses.  In relation to the field of quantum key distribution, our results confirm the gain-switched laser diode as a suitable light source, capable of providing phase-randomized coherent pulses at a clock rate of up to 2.5~GHz.
\end{abstract}

\maketitle

The development of high bit rate random number generators (RNGs) has recently attracted a lot of research interest.\cite{uchida08,kanter10,argyris10, williams10, qi10,jofre11,tang13} Random numbers are a vital resource for numerous applications ranging from cryptography to scientific simulation.
The quality of randomness has a direct impact  on the performance of the application.
Using cryptography as an example, the security of a cryptographic service can be significantly degraded by predictability within the random number sequence.  Some applications, such as quantum key distribution (QKD), require an ultrafast real-time feed of random numbers at a rate on the order of 1--10~Gb/s.\cite{honjo09}

The source of randomness must be physical.
Whereas pseudo-RNGs based on computer algorithms will eventually repeat themselves, physical RNGs rely on unpredictable outcomes of physical measurements.
For generators based on quantum mechanics  unpredictability can be derived rigorously from first principles.
Quantum random number generators (QRNG’s) relied upon single photon detection and offered limited bit rates.\cite{dynes08,furst10}
 The bit rate has now been increased by detecting macroscopic states instead.\cite{qi10, guo10, jofre11, williams10}  A recent report of a QRNG running at 12.5~Gb/s using amplified spontaneous emission is a notable example.\cite{williams10}
 At this speed QRNGs are now competitive in bit rate with other physical, non-quantum RNG’s, such as those based on chaotic lasers.\cite{uchida08, kanter10, argyris10}


It is desirable for a QRNG to operate with flexible clock frequencies and to be robust against fluctuations to ease its integration into a larger system.
Existing fast QRNGs, however, are often operated at uncommon clock frequencies with little prospect of frequency tunability.\cite{jofre11,tang13,abellan14}
Here, we demonstrate an interferometric RNG which tolerates $\pm$20\% deviation to its central operation clock frequency.
The key idea is to rely on the steady-state emission of long laser pulses of a gain-switched laser.
The long duration alleviates the requirements for temporal overlap and it also enhances the visibility due to the accompanying narrower spectral width.  At the same time, switching the gain above and below threshold guarantees the randomness for the electromagnetic phase of the generated optical pulses.
 We use a synchronous 8-bit digitizer and practical post-processing to achieve a random bit rate of up to 20~Gb/s that has passed stringent statistical tests for randomness.  \textcolor{black}{Furthermore, we examine the prospects of an even faster RNG with phase correlated laser pulses.} Finally, we discuss the implications of our findings to the field of QKD.

Spontaneous emission has been identified as a useful mechanism to generate quantum randomness, as it can be ascribed to the vacuum fluctuations of the optical field.\cite{williams10, qi10, jofre11}
\textcolor{black}{In a gain-switched laser diode, spontaneous emission affects pulse generation in two ways. Firstly, it affects the switch-on time of the laser pulse, introducing a random time-jitter of 2 to 20 ps.\cite{spano88}
Secondly, it influences the electromagnetic phase of each generated pulse. When the laser cavity prior to lasing is \textit{empty}, \textit{i.e.}, in the vacuum state, the lasing action is triggered entirely by spontaneous emission.
Spontaneous emission inherits its electromagnetic phase from the vacuum, the phase of which is totally unbiased and random. 
The empty cavity condition can be reached  when cavity photons have a sufficient time to decay prior to each lasing event.  Using realistic parameters, our calculation shows an average residual photon number of  $10^{-10}$  when a laser diode is gain-switched with a 2.5~GHz square wave (see Table I).} Phase randomness is readily converted into directly measurable intensity fluctuation, using an asymmetric Mach-Zehnder interferometer (AMZI).

\begin{figure}[t]
\centering
\includegraphics[width=1\columnwidth]{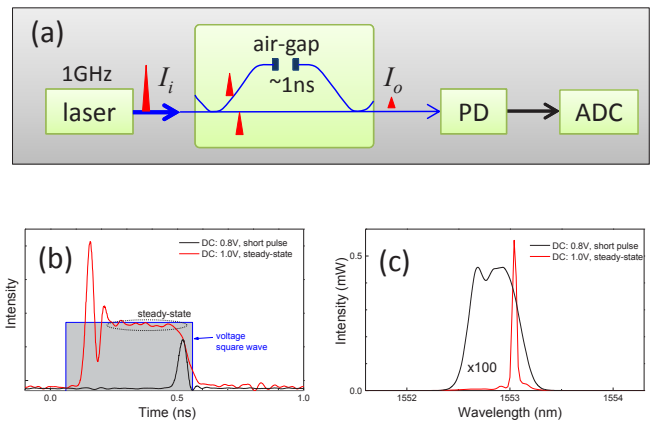}
\caption{ (a) Setup for random number generation; LD: pulsed laser diode;
PD: photodiode; and ADC: 8-bit analogue-digital converter.  (b) Temporal
profiles and (c) spectra  of laser emission with different DC biases. Timing of the square wave is approximately shown in (b).}
\label{fig:fig1}
\end{figure}

Fig.~\ref{fig:fig1}(a) shows the experimental setup.  It contains a gain-switched laser diode (bandwidth 10GHz), a fiber-optic AMZI, a photodiode (PD) and an 8-bit analogue-to-digital converter (ADC) of 13~GHz bandwidth.   The tunable delay line (air-gap) here is for experimental convenience only.   Through matching the interferometer differential delay to the laser clock frequency, interference occurs between optical pules emitted at different clock cycles in the output 50/50 beam splitter. All devices operate without temperature stabilisation.   The AMZI  drifts in phase at a rate of about $2\pi$ per 10--100 seconds.
\textcolor{black}{This drift produces an additional phase difference of  $<10^{-9}$ rad between any interfering pulse pairs,  which does not affect the ADC readout because of its finite resolution.}

Gain-switched laser diodes have been widely used for generating short pulses.\cite{pataca97, gobby04, dynes12}   As shown in Fig.~\ref{fig:fig1}(b), the laser diode emits short pulses of 37~ps full width at half maximum when driven by a 2~V, 1~GHz square wave superimposed upon a direct current (DC) bias of 0.8~V.  Short pulses are a result of the gradual build-up of carriers towards the lasing threshold.\cite{pataca97}
Frequency chirping has significantly broadened the emission spectrum, as shown in Fig.~\ref{fig:fig1}(c), due to the violent change in the carrier density and hence the refractive index of the laser cavity medium. These short pulses are not ideal for interferometric RNG. Firstly, their short duration demands a precise match of the AMZI differential delay to the laser clock frequency, leading to inflexible operation conditions.\cite{jofre11,tang13}   Secondly, the interference quality degrades significantly due to timing jitter and frequency chirp,  as described later.  And thirdly, short pulses demand precise synchronisation with the digitization electronics, adding further complexity to an RNG realization.

All the above disadvantages disappear when driving the laser into steady-state.  Raising the DC bias to 1.0~V,  lasing starts much earlier due to faster carrier built-up, as shown in Fig.~\ref{fig:fig1}(b).  The intensity initially oscillates, but relaxes rapidly into a steady-state after $\sim$100~ps.  The steady-state is a result of an approximate equilibrium between electrical injection and radiative depletion of charge carriers.  The equilibrium allows a stable refractive index of the laser cavity medium, and therefore a narrow spectrum.
 Its wavelength spectrum consists of a sharp, intense feature at 1553.04~nm, as shown in Fig.~\ref{fig:fig1}(c),  illustrating that almost no frequency chirp exists.

\begin{figure}[b]
\centering
\includegraphics[width=1\columnwidth]{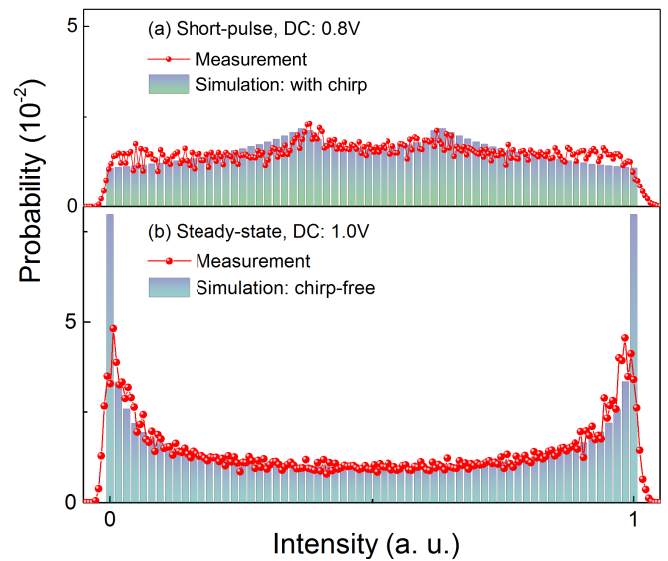}
\caption{ Experimental and simulated intensity distributions for the interference of (a) short pulses and  (b) steady-state emissions.   In simulation, we use random phase distribution,   and adopt 6-bit vertical resolution to reflect the noise in the 8-bit ADC.}
\label{fig:fig2}
\end{figure}

The different driving conditions have a critical influence on the interference properties.
In Fig.~\ref{fig:fig2} we compare short and steady-state pulse emissions using measured and simulated intensity distributions. The intensity distribution arises from interference of consecutive light pulses inside the AMZI. High (low) intensity corresponds to constructive (destructive) interference, respectively.
The AMZI differential delay is tuned to match the laser clock frequency exactly.  The ADC is synchronized to digitize the interference output signal at a sampling rate of 1~GHz, and then these samples are sorted to obtain the intensity distributions.   For both short and steady-state emission the intensity varies over the full range, however, the distributions show distinct difference. Whereas short pulse emission produces a rather uniform distribution, steady-state emission leads to sharp peaks at both ends of the sampling range.  These sharp peaks indicate high quality interference, as explained below.

We attribute the distribution difference observed in Fig.~\ref{fig:fig2} to time jitter ($\tau$) and frequency chirp ($\beta$) in the laser emission.   When interfering two coherent pulses of constant phase difference $\Delta \phi$, the output intensity $I_0$ is proportional to $(1+\cos \Delta \phi)$.  A uniform distribution of $\Delta \phi$ would produce two lateral peaks at both ends of the intensity distribution, as in Fig.~\ref{fig:fig2}(b).
However, the frequency chirp may prevent a constant phase difference between interfering pulses. When there is an arrival time difference $\Delta t$,  a linear frequency chirp generates a phase difference proportional to $\beta \Delta t \cdot t$, thus deteriorating the interference.   Only when two pulses are occasionally aligned ($\Delta t = 0$),   complete constructive (destructive) interference can take place.   This explains why complete interferences are observable in both Figs.~2(a) and (b).   With negligible chirp ($\beta \approx 0$), as in the case of steady-state emission,   the interference pulses always undergo identical phase evolution. Hence, complete interference is allowed even with temporal misalignment ($\Delta t \neq 0$).  This explains the enhanced probabilities at the destructive/constructive interferences in Fig.~\ref{fig:fig2}(b). We have simulated the interference intensity distributions using the following equation:
\begin{widetext}
\begin{equation}
\label{eq:eq1}
I_o =\frac{I_i}{2}\{1+\cos(\Delta \phi) \mathrm{sinc} (2\beta \Delta t T)+\sin(\Delta \phi) [1-\cos(2\beta \Delta t T)]/2\beta \Delta t T \},
\end{equation}
\end{widetext}
\noindent where  $I_i$ is the emitted pulse intensity, and $T$ is the detector sampling window duration.  We use $\beta \tau T = 2.5$ and  $\beta =0 $ (chirp-free) for the short pulses and steady-state emission, respectively.  The simulation is in excellent agreement with the experimental data, as shown in Fig.~\ref{fig:fig2}.

\begin{figure}[t]
\centering
\includegraphics[width=\columnwidth]{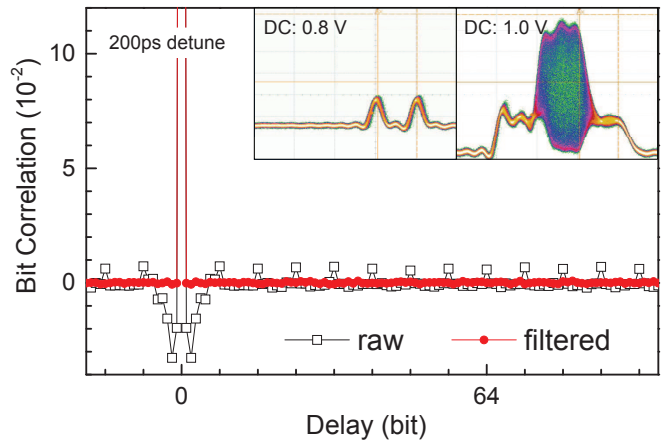}
\caption{Bit correlations before and after filtering  for RNG  with the AMZI 200~ps detuned to the 1~GHz laser clock rate. Inset: waveforms recorded by an oscilloscope in persistent display mode for  short pulses (left) and steady-state emissions (right) passing through the detuned AMZI.}
\label{fig:fig3}
\end{figure}

The long pulse duration now permits a misalignment of the differential delay of the AMZI without degrading the interference quality.  To demonstrate this,  we deliberately de-tune the AMZI delay by 20\%, \textit{i.e.},  we introduce an offset of 200~ps against the laser clock period of 1~ns. This temporal misalignment corresponds to a fiber length of \textcolor{black}{$\sim$40~mm}, easily achievable without an adjustable delay element, such as an air-gap.   The inset of Fig.~\ref{fig:fig3} compares the interferometer outputs, between short pulses and steady-state emissions, recorded by a real-time oscilloscope in persistent display mode.  With the present detuning of 200~ps, short pulses do not overlap after the interferometer and thus do not interfere.  In contrast, there is considerable temporal overlap achieved with steady-state emissions.   The duration of the temporal overlap is tunable by varying the DC bias.   In the overlapped part of the waveform, the recorded output fluctuates strongly between constructive and destructive interferences. Denser distribution at top and bottom intensities  suggest little degradation in interference quality.

We synchronise the ADC to sample the interfering part of the waveform and take 8-bits from each ADC sample as the RNG raw output. Analysis of these raw bits has revealed that bit correlation\cite{williams10} exists, and oscillates at a repetition period of 8-bits with a small, but statistically significant,  amplitude of $\sim$ 0.01, as shown in Fig.~\ref{fig:fig3}.  This oscillation arises mainly from the imperfect match of the ADC range to the PD signal output.  \textcolor{black}{In our case, the peak-to-peak amplitude of the PD signal is slightly less than the range of the ADC.
To confirm that the correlation does indeed stem from sampling mismatch, we have also computed the intensity autocorrelation (not shown), which does not show any oscillation.} 

\begin{table*}[]
\caption{Randomness test results of RNG at various rates of 8, 20 and 80~Gb/s using the NIST test suite.\cite{nist}   \textcolor{black}{For tests producing multiple $P$-values, a Kolmogorov-Smirnov [KS] uniformity test has been performed. For multiple proportion values, mean values are displayed.} Respective laser characteristics and AMZI detunings are also shown. The residual cavity photons are calculated for lasers driven electrically by square waves. }
\begin{tabular}[c]{|c||c|c|c|}\hline
\textbf{Laser repetition rate }(GHz) & $\mathbf{1}$ & $\mathbf{2.5}$ & $\mathbf{10}$\\\hline
Laser mode & steady-state  & steady-state & short pulses \\\hline
AMZI detuning (ps) & 200  & 80 & 0 \\ \hline
Random bit rate (Gb/s) & $8$ & $20$ & $80$\\\hline
Post-processing & \multicolumn{3} {c|}{FIR, $M=2$, $b_{\{0,1,2\}}=\{1, 2, 1\}$ }\\\hline
Residual cavity photons & $<10^{-35}$ & $<10^{-10}$ & $\sim10^2$\\\hline
Phase ripples & not visible & not visible & visible\\\hline\hline
\textbf{NIST test} & \multicolumn{3}{c|}{$P$-value / proportion}\\\hline
Frequency & $0.9558$ / $0.995$ & $0.9839$ / $0.990$ & $0.9323$ / $0.991$\\\hline
Block Frequency & $0.6870$ / $0.994$ & $0.8862$ / $0.993$ & $0.1056$ / $0.991$\\\hline
Cumulative Sums $^{\text{[KS]}}$ & $0.8617$ / $0.992$ & $0.6173$ / $0.991$ &  $0.8570$ / $0.992$\\\hline
Runs & $0.9558$ / $0.996$ & $0.7479$ / $0.985$ & $0.3925$
/ $0.989$\\\hline
Longest Run & $0.4136$ / $0.989$ & $0.3537$ / $0.992$ & $0.2544$ / $0.990$\\\hline
Rank & $0.0898$ / $0.994$ & $0.9756$ / $0.991$ & $0.6101$
/ $0.992$\\\hline
FFT & $0.7637$ / $0.994$ & $0.7459$ / $0.989$ & $0.9493$ / $0.987$\\\hline
Non Overlapping Template $^{\text{[KS]}}$ & $0.2832$ / $0.990$ & $0.5180$ /
$0.990$ & $0.1670$ / $0.989$\\\hline
Overlapping Template & $0.1728$ / $0.988$ & $0.6517$ / $0.989$ & $0.1319$ / $0.986$\\\hline
Universal & $0.6080$ / $0.989$ & $0.5914$ / $0.989$ & $0.0835$ / $0.993$\\\hline
Approximate Entropy & $0.8111$ / $0.991$ & $0.9473$ / $0.990$ & $0.7198$ / $0.990$\\\hline
Random Excursions $^{\text{[KS]}}$ & $0.4676$ / $0.991$ & $0.1763$ / $0.990$ & $0.8516$ / $0.988$\\\hline
Random Excursions Variant $^{\text{[KS]}}$ & $0.3590$ / $0.988$ & $0.0887$ /
$0.990$ & $0.1136$ / $0.993$\\\hline
Serial $^{\text{[KS]}}$ & $0.5948$ / $0.993$ & $0.8344$ / $0.993$ & $0.4863$ / $0.992$\\\hline
Linear Complexity & $0.5382$ / $0.990$ & $0.7399$ / $0.989$ & $0.8771$ / $0.993$\\\hline
\textbf{Result} & \textbf{Success} & \textbf{Success} &  \textbf{Success}\\\hline
\end{tabular}
\end{table*}

Any small but statistically significant correlations has to be removed by data post-processing. For that, we choose a finite-impulse response filter\cite{FIR} (FIR) to process the RNG raw output.
Its function here is to convert  the RNG raw output, \textit{i.e.}, a stream of 8-bit integers ($x\mathrm{[}n\mathrm{]}$),  into $y\mathrm{[}n\mathrm{]}$  using:
\begin{equation}\label{eq}
y[n]=\sum_{i=0}^{M}b_ix[n-i] \; \textrm{mod} \; 2^8,
\end{equation}
\noindent where $M \geq 1$ is an integer and $b_i$ are the filter coefficients. For simplicity, we choose $b_i = \frac{M!}{i!(M-i)!}$.  This choice is simply an $M$-th order addition of neighbouring elements. The operation  fuses bits of differing significances, thus achieving de-correlation of the  raw data.  Other coefficients can also be chosen readily.   For example, arbitrary sign can be chosen for each  $b_i$,  as the interference output is intrinsically non-deterministic.  We acknowledge that the derivative method used in the chaotic-laser based RNG's\cite{kanter10} is also a special case of FIR, in which $b_i = (-1)^i\frac{M!}{i!(M-i)!}$.
Minimum value for $M$ depends on the ADC filling factor.
As the interference produces full swing in intensity, $M=2$ is sufficient for de-correlation in our case.  The FIR effectively removes bit correlations as shown in Fig. 3 with the residue well within statistical fluctuation.
We notice that additional post-processing can be concatenated to the FIR in our scheme, for example, a randomness extractor based on the RNG quantum mechanical description.\cite{frauchiger13,law14} This can make our RNG  completely unpredictable even in principle, the output being true random numbers.

We collect a total of 1000 $\times$ 1~Mbits of data at a rate of 8~Gb/s. These random numbers are subjected to the National Institute of Standards and Technology (NIST) 800-22 statistical test suite.\cite{nist}   The test suite comprises 15 tests which result in a ``$p$-value".  The $p$-value for each test is defined as ``the probability a perfect RNG would have produced a sequence less random than the sequence that was tested, given the kind of non-randomness assessed by the test".\cite{nist}  A significance level for the $p$-value of $\alpha$ = 0.01 is selected. If the $p>\alpha$ for a particular test, the test is deemed to have been passed.  The ``$P$-value", \textit{i.e.}, the uniformity of the $p$-values, should be greater than 0.0001.  Finally, since we are testing a finite sequence, we expect some failure probability of passing statistically.
A parameter reflecting this probability is given by the proportion of passes.  For the RNG data analyzed in this paper, the passing proportion should be greater than 0.980.  As summarised in Table I, the RNG at a laser repetition rate of 1~GHz and a RNG rate of 8~Gb/s has passed the statistical test for randomness.

We can increase the random bit generation rate by clocking the laser at a higher repetition rate.  Driving it at 2.5~GHz, the duration of the steady-state emission shrinks, but remains sufficient for operation with 20\% detuning in the AMZI, \textit{i.e.}, 80~ps. At this clock frequency we obtain an RNG rate of 20~Gb/s (see Table I). This bit rate is sufficient to provide real-time feed to the most demanding applications, including GHz-clocked QKD.\cite{dynes12}

\begin{figure}[b]
\centering
\includegraphics[width=\columnwidth]{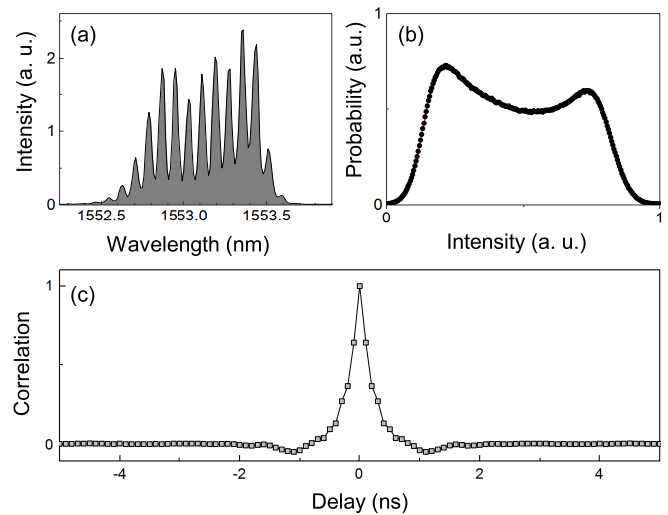}
\caption{(a) Laser spectrum with 10~GHz driving frequency; (b) Interference output histogram and  (c) intensity correlation. The AMZI is tuned to exactly 1~ns for interfering laser pulses 10 clocks apart.}
\label{fig:fig4}
\end{figure}

Further increasing the clock rate risks violation of the empty cavity condition. At 5~GHz frequency, we estimate a residual photon number of 0.01. Phase ripples already start to appear in the laser spectrum (not shown). When the frequency increases to 10~GHz, phase ripples dominate the laser spectrum, as shown in Fig.~\ref{fig:fig4}(a), with a constant spacing of 0.08~nm,  corresponding to the exact driving frequency.
We interfere 10~GHz laser pulses with a 1~ns AMZI.  While a uniform phase distribution is still suggested by the probability peaking in Fig.~\ref{fig:fig4}(b),  the intensity autocorrelation (Fig.~\ref{fig:fig4}(c)) has clearly identified that the electromagnetic phases are highly correlated  among pulses separated by 1.6~ns or less. The correlation is as high as 0.64 between adjacent clock cycles.
The level of correlation decreases with delay, and vanishes for delays greater than 2~ns.  We attribute this phase de-correlation to disturbance by spontaneous emission.   Ultrafast, seemingly random bits are possible with these optical pulses.
For a sampling rate of 10~GSamples/s, we obtain a bit rate of 80~Gb/s which has passed the statistical tests (see Table I). However, care needs to be taken to separate the effect of the phase correlation from the data.

Finally, let us discuss additional implication of our results to QKD.  Attenuated gain-switched lasers have become the light source of choice in practical QKD, because an efficient single photon source is still elusive.  In most security proofs,  attenuated lasers are assumed to generate coherent states of light with randomly distributed electromagnetic phases.\cite{gottesman04}  This assumption is crucial for the decoy-state technique,\cite{lo05,wang05}  but has never been examined experimentally.  In Ref. [\onlinecite{dixon09}], gain-switched lasers exhibit a second-order degree of coherence equal to 1 with above-threshold excitation.  This, together with high visibility interferences shown in Fig.~\ref{fig:fig2},  suggests a coherent state emission.\cite{loudon2000quantum}
The said visibility and the close resemblance between the measured and simulated distribution (Fig.~\ref{fig:fig2}(a)) indicates the electromagnetic phase is uniformly distributed.  Additionally, absence of intensity correlation at moderate laser clock rates ($\leq$2.5~GHz, not shown), as well as the successful outcome of the statistical tests, supports the phase randomness assumption. Therefore a gain-switched laser qualifies as the practical source in QKD up to $\sim$2.5~GHz clock rate.

To conclude, we have demonstrated a robust interferometric RNG that tolerates large temporal detuning of the interfering laser pulses. With a simple FIR filtering technique, this RNG has passed the statistical tests for randomness at a bit rate of up to 20~Gb/s. This can serve as a basis for a more refined QRNG in which the amount of quantum noise is readily quantified.  We also demonstrate and discuss  a 80Gb/s generation rate obtained by interfering phase correlated laser pulses.
Finally, we relate our findings to QKD, showing that a gain-switched laser qualifies as a suitable light source, even for high speed systems.

\newpage


\begin{thebibliography}{25}%
\makeatletter
\providecommand \@ifxundefined [1]{%
 \@ifx{#1\undefined}
}%
\providecommand \@ifnum [1]{%
 \ifnum #1\expandafter \@firstoftwo
 \else \expandafter \@secondoftwo
 \fi
}%
\providecommand \@ifx [1]{%
 \ifx #1\expandafter \@firstoftwo
 \else \expandafter \@secondoftwo
 \fi
}%
\providecommand \natexlab [1]{#1}%
\providecommand \enquote  [1]{``#1''}%
\providecommand \bibnamefont  [1]{#1}%
\providecommand \bibfnamefont [1]{#1}%
\providecommand \citenamefont [1]{#1}%
\providecommand \href@noop [0]{\@secondoftwo}%
\providecommand \href [0]{\begingroup \@sanitize@url \@href}%
\providecommand \@href[1]{\@@startlink{#1}\@@href}%
\providecommand \@@href[1]{\endgroup#1\@@endlink}%
\providecommand \@sanitize@url [0]{\catcode `\\12\catcode `\$12\catcode
  `\&12\catcode `\#12\catcode `\^12\catcode `\_12\catcode `\%12\relax}%
\providecommand \@@startlink[1]{}%
\providecommand \@@endlink[0]{}%
\providecommand \url  [0]{\begingroup\@sanitize@url \@url }%
\providecommand \@url [1]{\endgroup\@href {#1}{\urlprefix }}%
\providecommand \urlprefix  [0]{URL }%
\providecommand \Eprint [0]{\href }%
\providecommand \doibase [0]{http://dx.doi.org/}%
\providecommand \selectlanguage [0]{\@gobble}%
\providecommand \bibinfo  [0]{\@secondoftwo}%
\providecommand \bibfield  [0]{\@secondoftwo}%
\providecommand \translation [1]{[#1]}%
\providecommand \BibitemOpen [0]{}%
\providecommand \bibitemStop [0]{}%
\providecommand \bibitemNoStop [0]{.\EOS\space}%
\providecommand \EOS [0]{\spacefactor3000\relax}%
\providecommand \BibitemShut  [1]{\csname bibitem#1\endcsname}%
\let\auto@bib@innerbib\@empty
\bibitem [{\citenamefont {Uchida}\ \emph {et~al.}(2008)\citenamefont {Uchida},
  \citenamefont {Amano}, \citenamefont {Inoue}, \citenamefont {Hirano},
  \citenamefont {Naito}, \citenamefont {Someya}, \citenamefont {Oowada},
  \citenamefont {Kurashige}, \citenamefont {Shiki}, \citenamefont {Yoshimori},
  \citenamefont {Yoshimura},\ and\ \citenamefont {Davis}}]{uchida08}%
  \BibitemOpen
  \bibfield  {author} {\bibinfo {author} {\bibfnamefont {A.}~\bibnamefont
  {Uchida}}, \bibinfo {author} {\bibfnamefont {K.}~\bibnamefont {Amano}},
  \bibinfo {author} {\bibfnamefont {M.}~\bibnamefont {Inoue}}, \bibinfo
  {author} {\bibfnamefont {K.}~\bibnamefont {Hirano}}, \bibinfo {author}
  {\bibfnamefont {S.}~\bibnamefont {Naito}}, \bibinfo {author} {\bibfnamefont
  {H.}~\bibnamefont {Someya}}, \bibinfo {author} {\bibfnamefont
  {I.}~\bibnamefont {Oowada}}, \bibinfo {author} {\bibfnamefont
  {T.}~\bibnamefont {Kurashige}}, \bibinfo {author} {\bibfnamefont
  {M.}~\bibnamefont {Shiki}}, \bibinfo {author} {\bibfnamefont
  {S.}~\bibnamefont {Yoshimori}}, \bibinfo {author} {\bibfnamefont
  {K.}~\bibnamefont {Yoshimura}}, \ and\ \bibinfo {author} {\bibfnamefont
  {P.}~\bibnamefont {Davis}},\ }\href@noop {} {\bibfield  {journal} {\bibinfo
  {journal} {Nature Photon.}\ }\textbf {\bibinfo {volume} {2}},\ \bibinfo
  {pages} {728} (\bibinfo {year} {2008})}\BibitemShut {NoStop}%
\bibitem [{\citenamefont {Kanter}\ \emph {et~al.}(2010)\citenamefont {Kanter},
  \citenamefont {Aviad}, \citenamefont {Reidler}, \citenamefont {Cohen},\ and\
  \citenamefont {Rosenbluh}}]{kanter10}%
  \BibitemOpen
  \bibfield  {author} {\bibinfo {author} {\bibfnamefont {I.}~\bibnamefont
  {Kanter}}, \bibinfo {author} {\bibfnamefont {Y.}~\bibnamefont {Aviad}},
  \bibinfo {author} {\bibfnamefont {I.}~\bibnamefont {Reidler}}, \bibinfo
  {author} {\bibfnamefont {E.}~\bibnamefont {Cohen}}, \ and\ \bibinfo {author}
  {\bibfnamefont {M.}~\bibnamefont {Rosenbluh}},\ }\href@noop {} {\bibfield
  {journal} {\bibinfo  {journal} {Nature Photon.}\ }\textbf {\bibinfo {volume}
  {4}},\ \bibinfo {pages} {58} (\bibinfo {year} {2010})}\BibitemShut {NoStop}%
\bibitem [{\citenamefont {Argyris}\ \emph {et~al.}(2010)\citenamefont
  {Argyris}, \citenamefont {Deligiannidis}, \citenamefont {Pikasis},
  \citenamefont {Bogris},\ and\ \citenamefont {Syvridis}}]{argyris10}%
  \BibitemOpen
  \bibfield  {author} {\bibinfo {author} {\bibfnamefont {A.}~\bibnamefont
  {Argyris}}, \bibinfo {author} {\bibfnamefont {S.}~\bibnamefont
  {Deligiannidis}}, \bibinfo {author} {\bibfnamefont {E.}~\bibnamefont
  {Pikasis}}, \bibinfo {author} {\bibfnamefont {A.}~\bibnamefont {Bogris}}, \
  and\ \bibinfo {author} {\bibfnamefont {D.}~\bibnamefont {Syvridis}},\
  }\href@noop {} {\bibfield  {journal} {\bibinfo  {journal} {Opt. Express}\
  }\textbf {\bibinfo {volume} {18}},\ \bibinfo {pages} {18763 } (\bibinfo
  {year} {2010})}\BibitemShut {NoStop}%
\bibitem [{\citenamefont {Williams}\ \emph {et~al.}(2010)\citenamefont
  {Williams}, \citenamefont {Salevan}, \citenamefont {Li}, \citenamefont
  {Roy},\ and\ \citenamefont {Murphy}}]{williams10}%
  \BibitemOpen
  \bibfield  {author} {\bibinfo {author} {\bibfnamefont {C.~R.~S.}\
  \bibnamefont {Williams}}, \bibinfo {author} {\bibfnamefont {J.~C.}\
  \bibnamefont {Salevan}}, \bibinfo {author} {\bibfnamefont {X.}~\bibnamefont
  {Li}}, \bibinfo {author} {\bibfnamefont {R.}~\bibnamefont {Roy}}, \ and\
  \bibinfo {author} {\bibfnamefont {T.~E.}\ \bibnamefont {Murphy}},\ }\href
  {\doibase 10.1364/OE.18.023584} {\bibfield  {journal} {\bibinfo  {journal}
  {Opt. Express}\ }\textbf {\bibinfo {volume} {18}},\ \bibinfo {pages} {23584}
  (\bibinfo {year} {2010})}\BibitemShut {NoStop}%
\bibitem [{\citenamefont {Qi}\ \emph {et~al.}(2010)\citenamefont {Qi},
  \citenamefont {Chi}, \citenamefont {Lo},\ and\ \citenamefont {Qian}}]{qi10}%
  \BibitemOpen
  \bibfield  {author} {\bibinfo {author} {\bibfnamefont {B.}~\bibnamefont
  {Qi}}, \bibinfo {author} {\bibfnamefont {Y.-M.}\ \bibnamefont {Chi}},
  \bibinfo {author} {\bibfnamefont {H.-K.}\ \bibnamefont {Lo}}, \ and\ \bibinfo
  {author} {\bibfnamefont {L.}~\bibnamefont {Qian}},\ }\href {\doibase
  10.1364/OL.35.000312} {\bibfield  {journal} {\bibinfo  {journal} {Opt.
  Lett.}\ }\textbf {\bibinfo {volume} {35}},\ \bibinfo {pages} {312} (\bibinfo
  {year} {2010})}\BibitemShut {NoStop}%
\bibitem [{\citenamefont {Jofre}\ \emph {et~al.}(2011)\citenamefont {Jofre},
  \citenamefont {Curty}, \citenamefont {Steinlechner}, \citenamefont {Anzolin},
  \citenamefont {Torres}, \citenamefont {Mitchell},\ and\ \citenamefont
  {Pruneri}}]{jofre11}%
  \BibitemOpen
  \bibfield  {author} {\bibinfo {author} {\bibfnamefont {M.}~\bibnamefont
  {Jofre}}, \bibinfo {author} {\bibfnamefont {M.}~\bibnamefont {Curty}},
  \bibinfo {author} {\bibfnamefont {F.}~\bibnamefont {Steinlechner}}, \bibinfo
  {author} {\bibfnamefont {G.}~\bibnamefont {Anzolin}}, \bibinfo {author}
  {\bibfnamefont {J.~P.}\ \bibnamefont {Torres}}, \bibinfo {author}
  {\bibfnamefont {M.~W.}\ \bibnamefont {Mitchell}}, \ and\ \bibinfo {author}
  {\bibfnamefont {V.}~\bibnamefont {Pruneri}},\ }\href {\doibase
  10.1364/OE.19.020665} {\bibfield  {journal} {\bibinfo  {journal} {Opt.
  Express}\ }\textbf {\bibinfo {volume} {19}},\ \bibinfo {pages} {20665}
  (\bibinfo {year} {2011})}\BibitemShut {NoStop}%
\bibitem [{\citenamefont {Tang}\ \emph {et~al.}(2013)\citenamefont {Tang},
  \citenamefont {Jiang}, \citenamefont {Sun}, \citenamefont {Ma}, \citenamefont
  {Li},\ and\ \citenamefont {Liang}}]{tang13}%
  \BibitemOpen
  \bibfield  {author} {\bibinfo {author} {\bibfnamefont {G.~Z.}\ \bibnamefont
  {Tang}}, \bibinfo {author} {\bibfnamefont {M.~S.}\ \bibnamefont {Jiang}},
  \bibinfo {author} {\bibfnamefont {S.~H.}\ \bibnamefont {Sun}}, \bibinfo
  {author} {\bibfnamefont {X.~C.}\ \bibnamefont {Ma}}, \bibinfo {author}
  {\bibfnamefont {C.~Y.}\ \bibnamefont {Li}}, \ and\ \bibinfo {author}
  {\bibfnamefont {L.~M.}\ \bibnamefont {Liang}},\ }\href@noop {} {\bibfield
  {journal} {\bibinfo  {journal} {Chin. Phys. Lett.}\ }\textbf {\bibinfo
  {volume} {30}},\ \bibinfo {pages} {114207} (\bibinfo {year}
  {2013})}\BibitemShut {NoStop}%
\bibitem [{\citenamefont {Honjo}\ \emph {et~al.}(2009)\citenamefont {Honjo},
  \citenamefont {Uchida}, \citenamefont {Amano}, \citenamefont {Hirano},
  \citenamefont {Someya}, \citenamefont {Okumura}, \citenamefont {Yoshimura},
  \citenamefont {Davis},\ and\ \citenamefont {Tokura}}]{honjo09}%
  \BibitemOpen
  \bibfield  {author} {\bibinfo {author} {\bibfnamefont {T.}~\bibnamefont
  {Honjo}}, \bibinfo {author} {\bibfnamefont {A.}~\bibnamefont {Uchida}},
  \bibinfo {author} {\bibfnamefont {K.}~\bibnamefont {Amano}}, \bibinfo
  {author} {\bibfnamefont {K.}~\bibnamefont {Hirano}}, \bibinfo {author}
  {\bibfnamefont {H.}~\bibnamefont {Someya}}, \bibinfo {author} {\bibfnamefont
  {H.}~\bibnamefont {Okumura}}, \bibinfo {author} {\bibfnamefont
  {K.}~\bibnamefont {Yoshimura}}, \bibinfo {author} {\bibfnamefont
  {P.}~\bibnamefont {Davis}}, \ and\ \bibinfo {author} {\bibfnamefont
  {Y.}~\bibnamefont {Tokura}},\ }\href {\doibase 10.1364/OE.17.009053}
  {\bibfield  {journal} {\bibinfo  {journal} {Opt. Express}\ }\textbf {\bibinfo
  {volume} {17}},\ \bibinfo {pages} {9053} (\bibinfo {year}
  {2009})}\BibitemShut {NoStop}%
\bibitem [{\citenamefont {Dynes}\ \emph {et~al.}(2008)\citenamefont {Dynes},
  \citenamefont {Yuan}, \citenamefont {Sharpe},\ and\ \citenamefont
  {Shields}}]{dynes08}%
  \BibitemOpen
  \bibfield  {author} {\bibinfo {author} {\bibfnamefont {J.~F.}\ \bibnamefont
  {Dynes}}, \bibinfo {author} {\bibfnamefont {Z.~L.}\ \bibnamefont {Yuan}},
  \bibinfo {author} {\bibfnamefont {A.~W.}\ \bibnamefont {Sharpe}}, \ and\
  \bibinfo {author} {\bibfnamefont {A.~J.}\ \bibnamefont {Shields}},\
  }\href@noop {} {\bibfield  {journal} {\bibinfo  {journal} {Appl. Phys.
  Lett.}\ }\textbf {\bibinfo {volume} {93}},\ \bibinfo {eid} {031109} (\bibinfo
  {year} {2008})}\BibitemShut {NoStop}%
\bibitem [{\citenamefont {F\"urst}\ \emph {et~al.}(2010)\citenamefont
  {F\"urst}, \citenamefont {Weier}, \citenamefont {Nauerth}, \citenamefont
  {Marangon}, \citenamefont {Kurtsiefer},\ and\ \citenamefont
  {Weinfurter}}]{furst10}%
  \BibitemOpen
  \bibfield  {author} {\bibinfo {author} {\bibfnamefont {M.}~\bibnamefont
  {F\"urst}}, \bibinfo {author} {\bibfnamefont {H.}~\bibnamefont {Weier}},
  \bibinfo {author} {\bibfnamefont {S.}~\bibnamefont {Nauerth}}, \bibinfo
  {author} {\bibfnamefont {D.~G.}\ \bibnamefont {Marangon}}, \bibinfo {author}
  {\bibfnamefont {C.}~\bibnamefont {Kurtsiefer}}, \ and\ \bibinfo {author}
  {\bibfnamefont {H.}~\bibnamefont {Weinfurter}},\ }\href@noop {} {\bibfield
  {journal} {\bibinfo  {journal} {Opt. Express}\ }\textbf {\bibinfo {volume}
  {18}},\ \bibinfo {pages} {13029} (\bibinfo {year} {2010})}\BibitemShut
  {NoStop}%
\bibitem [{\citenamefont {Guo}\ \emph {et~al.}(2010)\citenamefont {Guo},
  \citenamefont {Tang}, \citenamefont {Liu},\ and\ \citenamefont
  {Wei}}]{guo10}%
  \BibitemOpen
  \bibfield  {author} {\bibinfo {author} {\bibfnamefont {H.}~\bibnamefont
  {Guo}}, \bibinfo {author} {\bibfnamefont {W.}~\bibnamefont {Tang}}, \bibinfo
  {author} {\bibfnamefont {Y.}~\bibnamefont {Liu}}, \ and\ \bibinfo {author}
  {\bibfnamefont {W.}~\bibnamefont {Wei}},\ }\href {\doibase
  10.1103/PhysRevE.81.051137} {\bibfield  {journal} {\bibinfo  {journal} {Phys.
  Rev. E}\ }\textbf {\bibinfo {volume} {81}},\ \bibinfo {pages} {051137}
  (\bibinfo {year} {2010})}\BibitemShut {NoStop}%
\bibitem [{\citenamefont {Abell\'{a}n}\ \emph {et~al.}(2014)\citenamefont
  {Abell\'{a}n}, \citenamefont {Amaya}, \citenamefont {Jofre}, \citenamefont
  {Curty}, \citenamefont {Ac\'{i}n}, \citenamefont {Capmany}, \citenamefont
  {Pruneri},\ and\ \citenamefont {Mitchell}}]{abellan14}%
  \BibitemOpen
  \bibfield  {author} {\bibinfo {author} {\bibfnamefont {C.}~\bibnamefont
  {Abell\'{a}n}}, \bibinfo {author} {\bibfnamefont {W.}~\bibnamefont {Amaya}},
  \bibinfo {author} {\bibfnamefont {M.}~\bibnamefont {Jofre}}, \bibinfo
  {author} {\bibfnamefont {M.}~\bibnamefont {Curty}}, \bibinfo {author}
  {\bibfnamefont {A.}~\bibnamefont {Ac\'{i}n}}, \bibinfo {author}
  {\bibfnamefont {J.}~\bibnamefont {Capmany}}, \bibinfo {author} {\bibfnamefont
  {V.}~\bibnamefont {Pruneri}}, \ and\ \bibinfo {author} {\bibfnamefont
  {M.~W.}\ \bibnamefont {Mitchell}},\ }\href {\doibase 10.1364/OE.22.001645}
  {\bibfield  {journal} {\bibinfo  {journal} {Opt. Express}\ }\textbf {\bibinfo
  {volume} {22}},\ \bibinfo {pages} {1645} (\bibinfo {year}
  {2014})}\BibitemShut {NoStop}%
\bibitem [{\citenamefont {Spano}\ \emph {et~al.}(1988)\citenamefont {Spano},
  \citenamefont {D'Ottavi}, \citenamefont {Mecozzi},\ and\ \citenamefont
  {Daino}}]{spano88}%
  \BibitemOpen
  \bibfield  {author} {\bibinfo {author} {\bibfnamefont {P.}~\bibnamefont
  {Spano}}, \bibinfo {author} {\bibfnamefont {A.}~\bibnamefont {D'Ottavi}},
  \bibinfo {author} {\bibfnamefont {A.}~\bibnamefont {Mecozzi}}, \ and\
  \bibinfo {author} {\bibfnamefont {B.}~\bibnamefont {Daino}},\ }\href@noop {}
  {\bibfield  {journal} {\bibinfo  {journal} {Appl. Phys. Lett.}\ }\textbf
  {\bibinfo {volume} {52}},\ \bibinfo {pages} {2203} (\bibinfo {year}
  {1988})}\BibitemShut {NoStop}%
\bibitem [{\citenamefont {Pataca}\ \emph {et~al.}(1997)\citenamefont {Pataca},
  \citenamefont {Gunning}, \citenamefont {Rocha}, \citenamefont {Lucek},
  \citenamefont {Kashyap}, \citenamefont {Smith}, \citenamefont {Moodie},
  \citenamefont {Davey}, \citenamefont {Souza},\ and\ \citenamefont
  {Siddiqui}}]{pataca97}%
  \BibitemOpen
  \bibfield  {author} {\bibinfo {author} {\bibfnamefont {D.~M.}\ \bibnamefont
  {Pataca}}, \bibinfo {author} {\bibfnamefont {P.}~\bibnamefont {Gunning}},
  \bibinfo {author} {\bibfnamefont {M.~L.}\ \bibnamefont {Rocha}}, \bibinfo
  {author} {\bibfnamefont {J.~K.}\ \bibnamefont {Lucek}}, \bibinfo {author}
  {\bibfnamefont {R.}~\bibnamefont {Kashyap}}, \bibinfo {author} {\bibfnamefont
  {K.}~\bibnamefont {Smith}}, \bibinfo {author} {\bibfnamefont {D.~G.}\
  \bibnamefont {Moodie}}, \bibinfo {author} {\bibfnamefont {R.~P.}\
  \bibnamefont {Davey}}, \bibinfo {author} {\bibfnamefont {R.~F.}\ \bibnamefont
  {Souza}}, \ and\ \bibinfo {author} {\bibfnamefont {A.~S.}\ \bibnamefont
  {Siddiqui}},\ }\href@noop {} {\bibfield  {journal} {\bibinfo  {journal} {J.
  Microwav. Optoelectron.}\ }\textbf {\bibinfo {volume} {1}},\ \bibinfo {pages}
  {46} (\bibinfo {year} {1997})}\BibitemShut {NoStop}%
\bibitem [{\citenamefont {Gobby}\ \emph {et~al.}(2004)\citenamefont {Gobby},
  \citenamefont {Yuan},\ and\ \citenamefont {Shields}}]{gobby04}%
  \BibitemOpen
  \bibfield  {author} {\bibinfo {author} {\bibfnamefont {C.}~\bibnamefont
  {Gobby}}, \bibinfo {author} {\bibfnamefont {Z.~L.}\ \bibnamefont {Yuan}}, \
  and\ \bibinfo {author} {\bibfnamefont {A.~J.}\ \bibnamefont {Shields}},\
  }\href@noop {} {\bibfield  {journal} {\bibinfo  {journal} {Appl. Phys.
  Lett.}\ }\textbf {\bibinfo {volume} {84}},\ \bibinfo {pages} {3762} (\bibinfo
  {year} {2004})}\BibitemShut {NoStop}%
\bibitem [{\citenamefont {Dynes}\ \emph {et~al.}(2012)\citenamefont {Dynes},
  \citenamefont {Choi}, \citenamefont {Sharpe}, \citenamefont {Dixon},
  \citenamefont {Yuan}, \citenamefont {Fujiwara}, \citenamefont {Sasaki},\ and\
  \citenamefont {Shields}}]{dynes12}%
  \BibitemOpen
  \bibfield  {author} {\bibinfo {author} {\bibfnamefont {J.~F.}\ \bibnamefont
  {Dynes}}, \bibinfo {author} {\bibfnamefont {I.}~\bibnamefont {Choi}},
  \bibinfo {author} {\bibfnamefont {A.~W.}\ \bibnamefont {Sharpe}}, \bibinfo
  {author} {\bibfnamefont {A.~R.}\ \bibnamefont {Dixon}}, \bibinfo {author}
  {\bibfnamefont {Z.~L.}\ \bibnamefont {Yuan}}, \bibinfo {author}
  {\bibfnamefont {M.}~\bibnamefont {Fujiwara}}, \bibinfo {author}
  {\bibfnamefont {M.}~\bibnamefont {Sasaki}}, \ and\ \bibinfo {author}
  {\bibfnamefont {A.~J.}\ \bibnamefont {Shields}},\ }\href@noop {} {\bibfield
  {journal} {\bibinfo  {journal} {Opt. Express}\ }\textbf {\bibinfo {volume}
  {20}},\ \bibinfo {pages} {16339} (\bibinfo {year} {2012})}\BibitemShut
  {NoStop}%
\bibitem [{\citenamefont {Ifeachor}\ and\ \citenamefont {Jervis}(2002)}]{FIR}%
  \BibitemOpen
  \bibfield  {author} {\bibinfo {author} {\bibfnamefont {E.~C.}\ \bibnamefont
  {Ifeachor}}\ and\ \bibinfo {author} {\bibfnamefont {B.~W.}\ \bibnamefont
  {Jervis}},\ }\href@noop {} {\emph {\bibinfo {title} {Digital signal
  processing: a practical approach}}}\ (\bibinfo  {publisher} {Pearson
  Education},\ \bibinfo {year} {2002})\BibitemShut {NoStop}%
\bibitem [{\citenamefont {Frauchiger}\ \emph {et~al.}(2013)\citenamefont
  {Frauchiger}, \citenamefont {Renner},\ and\ \citenamefont
  {Troyer}}]{frauchiger13}%
  \BibitemOpen
  \bibfield  {author} {\bibinfo {author} {\bibfnamefont {D.}~\bibnamefont
  {Frauchiger}}, \bibinfo {author} {\bibfnamefont {R.}~\bibnamefont {Renner}},
  \ and\ \bibinfo {author} {\bibfnamefont {M.}~\bibnamefont {Troyer}},\
  }\href@noop {} {\  (\bibinfo {year} {2013})},\ \Eprint
  {http://arxiv.org/abs/quant-ph/1311.4547v1} {quant-ph/1311.4547v1}
  \BibitemShut {NoStop}%
\bibitem [{\citenamefont {Law}\ \emph {et~al.}(2014)\citenamefont {Law},
  \citenamefont {Bancal},\ and\ \citenamefont {Scarani}}]{law14}%
  \BibitemOpen
  \bibfield  {author} {\bibinfo {author} {\bibfnamefont {Y.~Z.}\ \bibnamefont
  {Law}}, \bibinfo {author} {\bibfnamefont {J.-D.}\ \bibnamefont {Bancal}}, \
  and\ \bibinfo {author} {\bibfnamefont {V.}~\bibnamefont {Scarani}},\
  }\href@noop {} {\  (\bibinfo {year} {2014})},\ \Eprint
  {http://arxiv.org/abs/quant-ph/1401.4243v1} {quant-ph/1401.4243v1}
  \BibitemShut {NoStop}%
\bibitem [{\citenamefont {Rukhin}\ \emph {et~al.}(2001)\citenamefont {Rukhin},
  \citenamefont {Soto}, \citenamefont {Nechvatal}, \citenamefont {Smid},
  \citenamefont {Barker}, \citenamefont {Leigh}, \citenamefont {Levenson},
  \citenamefont {Vangel}, \citenamefont {Banks}, \citenamefont {Heckert} \emph
  {et~al.}}]{nist}%
  \BibitemOpen
  \bibfield  {author} {\bibinfo {author} {\bibfnamefont {A.}~\bibnamefont
  {Rukhin}}, \bibinfo {author} {\bibfnamefont {J.}~\bibnamefont {Soto}},
  \bibinfo {author} {\bibfnamefont {J.}~\bibnamefont {Nechvatal}}, \bibinfo
  {author} {\bibfnamefont {M.}~\bibnamefont {Smid}}, \bibinfo {author}
  {\bibfnamefont {E.}~\bibnamefont {Barker}}, \bibinfo {author} {\bibfnamefont
  {S.}~\bibnamefont {Leigh}}, \bibinfo {author} {\bibfnamefont
  {M.}~\bibnamefont {Levenson}}, \bibinfo {author} {\bibfnamefont
  {M.}~\bibnamefont {Vangel}}, \bibinfo {author} {\bibfnamefont
  {D.}~\bibnamefont {Banks}}, \bibinfo {author} {\bibfnamefont
  {A.}~\bibnamefont {Heckert}},  \emph {et~al.},\ }\href@noop {} {\bibfield
  {journal} {\bibinfo  {journal} {A statistical test suite for random and
  pseudorandom number generators for cryptographic applications}\ } (\bibinfo
  {year} {2001})}\BibitemShut {NoStop}%
\bibitem [{\citenamefont {Gottesman}\ \emph {et~al.}(2004)\citenamefont
  {Gottesman}, \citenamefont {Lo}, \citenamefont {L{\"u}tkenhaus},\ and\
  \citenamefont {Preskill}}]{gottesman04}%
  \BibitemOpen
  \bibfield  {author} {\bibinfo {author} {\bibfnamefont {D.}~\bibnamefont
  {Gottesman}}, \bibinfo {author} {\bibfnamefont {H.-K.}\ \bibnamefont {Lo}},
  \bibinfo {author} {\bibfnamefont {N.}~\bibnamefont {L{\"u}tkenhaus}}, \ and\
  \bibinfo {author} {\bibfnamefont {J.}~\bibnamefont {Preskill}},\ }\href@noop
  {} {\bibfield  {journal} {\bibinfo  {journal} {Quantum Information \&
  Computation}\ }\textbf {\bibinfo {volume} {4}},\ \bibinfo {pages} {325}
  (\bibinfo {year} {2004})}\BibitemShut {NoStop}%
\bibitem [{\citenamefont {Lo}\ \emph {et~al.}(2005)\citenamefont {Lo},
  \citenamefont {Ma},\ and\ \citenamefont {Chen}}]{lo05}%
  \BibitemOpen
  \bibfield  {author} {\bibinfo {author} {\bibfnamefont {H.~K.}\ \bibnamefont
  {Lo}}, \bibinfo {author} {\bibfnamefont {X.~F.}\ \bibnamefont {Ma}}, \ and\
  \bibinfo {author} {\bibfnamefont {K.}~\bibnamefont {Chen}},\ }\href@noop {}
  {\bibfield  {journal} {\bibinfo  {journal} {Phys. Rev. Lett.}\ }\textbf
  {\bibinfo {volume} {94}},\ \bibinfo {pages} {230504} (\bibinfo {year}
  {2005})}\BibitemShut {NoStop}%
\bibitem [{\citenamefont {Wang}(2005)}]{wang05}%
  \BibitemOpen
  \bibfield  {author} {\bibinfo {author} {\bibfnamefont {X.~B.}\ \bibnamefont
  {Wang}},\ }\href@noop {} {\bibfield  {journal} {\bibinfo  {journal} {Phys.
  Rev. Lett.}\ }\textbf {\bibinfo {volume} {94}},\ \bibinfo {eid} {230503}
  (\bibinfo {year} {2005})}\BibitemShut {NoStop}%
\bibitem [{\citenamefont {Dixon}\ \emph {et~al.}(2009)\citenamefont {Dixon},
  \citenamefont {Dynes}, \citenamefont {Yuan}, \citenamefont {Sharpe},
  \citenamefont {Bennett},\ and\ \citenamefont {Shields}}]{dixon09}%
  \BibitemOpen
  \bibfield  {author} {\bibinfo {author} {\bibfnamefont {A.~R.}\ \bibnamefont
  {Dixon}}, \bibinfo {author} {\bibfnamefont {J.~F.}\ \bibnamefont {Dynes}},
  \bibinfo {author} {\bibfnamefont {Z.~L.}\ \bibnamefont {Yuan}}, \bibinfo
  {author} {\bibfnamefont {A.~W.}\ \bibnamefont {Sharpe}}, \bibinfo {author}
  {\bibfnamefont {A.~J.}\ \bibnamefont {Bennett}}, \ and\ \bibinfo {author}
  {\bibfnamefont {A.~J.}\ \bibnamefont {Shields}},\ }\href@noop {} {\bibfield
  {journal} {\bibinfo  {journal} {Appl. Phys. Lett.}\ }\textbf {\bibinfo
  {volume} {94}},\ \bibinfo {pages} {231113} (\bibinfo {year}
  {2009})}\BibitemShut {NoStop}%
\bibitem [{\citenamefont {Loudon}(2000)}]{loudon2000quantum}%
  \BibitemOpen
  \bibfield  {author} {\bibinfo {author} {\bibfnamefont {R.}~\bibnamefont
  {Loudon}},\ }\href@noop {} {\emph {\bibinfo {title} {The quantum theory of
  light}}}\ (\bibinfo  {publisher} {Oxford university press},\ \bibinfo {year}
  {2000})\BibitemShut {NoStop}%
\end{thebibliography}
%

\newpage

\end{document}